\documentclass[12pt]{article}
\usepackage{times,graphicx}
\usepackage[super]{natbib} 
\usepackage{geometry}
\usepackage[utf8]{inputenc}
\usepackage{enumitem,amssymb}
\usepackage{ragged2e}
\newlist{thematic}{itemize}{8}
\setlist[thematic]{label=$\square$}
\usepackage{pifont}
\newcommand{\cmark}{\ding{51}}%
\newcommand{\done}{\rlap{$\square$}{\raisebox{2pt}{\large\hspace{1pt}\cmark}}%
\hspace{-2.5pt}}

\begin{document}
\begin{flushleft}
\huge
Astro2020 Science White Paper \linebreak

Populations of Black Holes in Binaries \linebreak
\normalsize

\noindent \textbf{Thematic Areas:} \hspace*{60pt} $\square$ Planetary Systems \hspace*{10pt} $\square$ Star and Planet Formation \hspace*{20pt}\linebreak
$\done$ Formation and Evolution of Compact Objects \hspace*{31pt} $\square$ Cosmology and Fundamental Physics \linebreak
  $\square$  Stars and Stellar Evolution \hspace*{1pt} $\square$ Resolved Stellar Populations and their Environments \hspace*{40pt} \linebreak
  $\square$    Galaxy Evolution   \hspace*{45pt} $\square$             Multi-Messenger Astronomy and Astrophysics \hspace*{65pt} \linebreak
  
\textbf{Principal Author:}

Name:	Thomas J. Maccarone
 \linebreak						
Institution:  Texas Tech University
 \linebreak
Email: thomas.maccarone@ttu.edu
 \linebreak
Phone:  (806) 834-3760
 \linebreak
 
\textbf{Co-authors:} (names and institutions)
  \linebreak
Laura Chomiuk (Michigan State University), James Miller-Jones (Curtin University),  Eric C. Bellm (University of Washington), Katelyn Breivik (CITA), Chris L. Fryer (Los Alamos National Laboratory), Vicky Kalogera (Northwestern University), Shane Larson (Northwestern), Jerome Orosz (San Diego State University),  James F. Steiner (MIT), Jay Strader (Michigan State University), John A. Tomsick (UC Berkeley)

\textbf{Abstract  (optional):}
Black holes in binary star systems are vital for understanding the process of producing gravitational wave sources, understanding how supernovae work, and for providing fossil evidence for the high mass stars from earlier in the Universe.  At the present time, sample sizes of these objects, and especially of black holes in binaries, are quite limited.  Furthermore, more precise measurements of the binary parameters are needed, as well.  With improvements primarily in X-ray and radio astronomy capabilities, it should be possible to build much larger samples of much better measured black hole binaries.
\end{flushleft}
\pagebreak
{\bf Overview: the importance of stellar compact object populations}

Understanding the populations of compact objects in binary systems is vital for a variety of reasons in astrophysics.  These objects are the progenitors of gravitational wave sources, and give one of the few means to understand the supernova process\citep{Belczynski12}.  Stellar remnants are also a key probe of the initial mass function, as one of the few indicators of the number of high mass stars that remain long after a burst of star formation \citep{peacock14}. 

{\it Binary evolution}\\
Binary evolution is important for producing gravitational wave sources, producing the r-process elements, through neutron star mergers, and iron, through Type Ia supernovae.  Close binaries are also responsible for maintaining the cores of globular clusters \citep{fregeau03}.  X-ray binaries are an ideal tracer of binary stellar evolution as their populations can be discovered even in relatively distant galaxies\citep{kundu07}.  This is all in addition to the fact that black holes are interesting in their own right.

{\it Understanding supernovae}\\
Supernovae  take place with their central engines embedded in opaque atmospheres.  Understanding the explosion mechanism then relies on gravitational waves and neutrinos which can escape the envelopes (but are likely to be detectable for very small numbers of events for the foreseeable future) or making use of the properties -- masses, spins and natal kicks -- of the stellar remnants.  

Already, this approach has shown its potential.  From the masses of known black holes and neutron stars in X-ray binaries, evidence has been found for a ``mass gap'' between the most massive neutron stars and the least massive black holes (Figure 1 left \citep{Ozel10,Farr11}).  If confirmed, this mass gap would imply that supernova explosions take place on timescales of 100-200 milliseconds\citep{Belczynski12}, contrary to the ``standard'' ideas about how supernova explosion mechanisms work.   Additionally, the LIGO discoveries, which are heavily weighted toward black holes larger than those typically seen in the Milky Way suggest that either the LIGO objects mostly form dynamically in globular clusters, or through a channel different from that for typical X-ray binaries' black holes.

\begin{figure}[!b]
    \centering
    \includegraphics[width=3in]{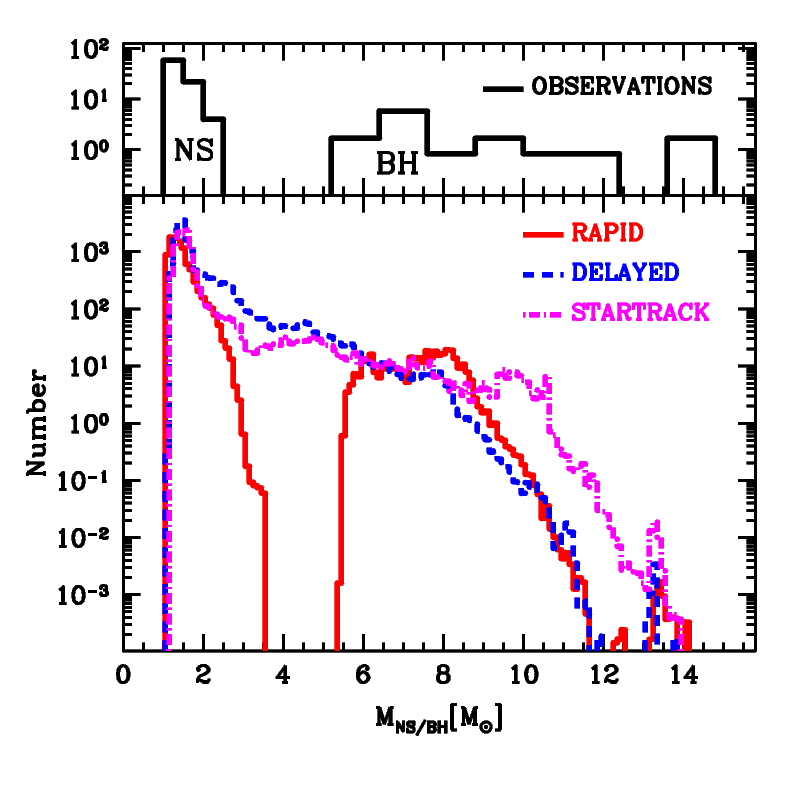}  \includegraphics[width=3in]{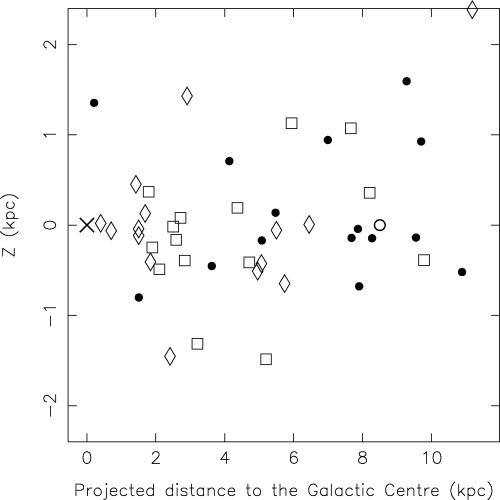}
    \caption{Left: The mass distribution of compact objects, indicating that a mass gap requires a rapid supernova explosion after the initial collapse begins\citep{Belczynski12}.  Right: The scale height distribution of X-ray binaries\citep{2004MNRAS.354..355J}, illustrating that large distances from the Galactic Plane are about as likely for black holes (filled symbols) as for neutron stars (open symbols).}
    \label{fig:my_label}
\end{figure}

The remnant mass distributions are presently collected primarily from X-ray binaries and binary pulsars.  Most of the accreting black holes are discovered in X-ray outbursts, and then have their masses estimated in their quiescent phases\citep{Casares14}.  Radial velocity amplitudes are measured using optical spectroscopic monitoring.  Mass ratios between the two binary components are estimated using rotational broadening of the donor star lines.  Orbital inclination angles are estimated using ellipsoidal modulations due to the distorted shape of the donor star.

Key systematics are likely to remain in the mass estimates, as even in deep ``quiescence'', most X-ray binaries have a non-trivial amount of light from the accretion disk and/or jet which contribute in the optical and infrared bands where the ellipsoidal modulation measurements are made, and the spectral shape of this emission is not well constrained because of uncertainties in the magnitudes of the jet contributions and in the temperature profiles of the outer parts of the accretion disks.  While, for Roche-lobe overflowing systems, the density of the donor star is fixed for a given orbital period, the donor stars in X-ray binaries are typically "bloated", with larger radii for a given spectral type than main sequence stars, but the level of bloating varies from system to system.  As a result, there is typically a degeneracy among the mass of the compact object, the distance to the system and the brightness of the accretion light.  This degeneracy can be broken with long, intensive monitoring campaigns over which the level of accretion light varies, since the donor star itself should not change its properties substantially on few year timescales\citep{Cantrell10}.  Alternatively, with good geometric parallax measurements, the distance could be fixed and the other parameters solved.  When donor stars are not Roche-lobe filling, solving the properties of the binary is more complicated, but can, in some cases still be done quite effectively by using the rotational properties of the binary, but these still benefit strongly from precise distance estimates\citep{Orosz11}.

Furthermore, in the future, we should endeavor to obtain substantial samples of objects from a variety of detection mechanisms.  Because X-ray binaries necessarily have common envelope evolution phases, the mass distribution of compact objects in them may not be representative of the mass distributions of compact objects formed without common envelopes.  The latter class of objects can be discovered from microlensing surveys\citep{Bennett2002} (although the precision on the mass estimates will then be rather poor without exquisite astrometry) or perhaps from accretion from the interstellar medium, in the radio band (but with very imprecise and model dependent mass estimates) \citep{Maccarone05,Fender2013}.  Instead, wide binaries represent the best hope for measurement of the mass distributions of black holes and neutron stars produced without common envelope phases.  It is expected that Gaia will discover many of these \citep{Barstow14,Mashian17, Breivik17}, and that, since they will be found from astrometric wobble, the inclination angles of the binaries will be directly measurable from the astrometry.  Furthermore, others can be discovered by radial velocity surveys\citep{Thompson18,Giesers18}.  

The other class of X-ray binaries that may form without common envelope phases are those in globular clusters.  In globular clusters, it is likely that most X-ray binaries form via some dynamical process\citep{Clark75}.  For pulsars in globular clusters, timing measurements often give good mass estimates.  For the black hole candidates in globular clusters, which are mostly either extragalactic objects selected on the basis of very high X-ray luminosities and variability\citep{Maccarone07,Maccarone11,Irwin16}, or Galactic objects selected on the basis of their radio emission\citep{2012Natur.490...71S,2013ApJ...777...69C,millerjones15}, mass estimates are challenging because crowding makes ground-based optical spectroscopy challenging.

{\bf Natal kicks}\\
The other key property of X-ray binaries is the natal kick distribution.  It has long been known that isolated pulsars show typical space velocities of $\sim$ 200 km/sec.  The scale height distributions of X-ray binaries are typically about 1 kpc, much larger than the scale height distribution of massive stars (e.g.\citep{2004MNRAS.354..355J}; Figure 1 right).  

Two main mechanisms may impart these large space velocities to compact objects:  the Blaauw mechanism\citep{Blaauw61}, and asymmetric natal kicks.  In the Blaauw mechanism, total momentum is conserved as a the supernova progenitor ejects material symmetrically about its rapidly orbiting core.  If the ejection mostly happens in much less than an orbital period, a recoil is applied to the binary to counteract the net loss of momentum of the ejecta.

The second mechanism is that of an asymmetric kick.  In many models of supernovae, various mechanisms can lead to asymmetries in the supernovae themselves.  These can apply kicks of arbitrary size and direction.  With large samples of objects with exquisite measurements, the two mechanisms can be disentangled, because the Blauuw mechanism will produce kicks confined to the plane of the binary system, while the asymmetric kicks can be in any direction.  Ideally, direct measurement of the binaries' position angles would be used to disentangle these effects, and can be done for the small number of systems with both resolved jet proper motions and astrometric wobble.  X-ray polarimetry might also be able to establish both the disk's position angle and the misalignment angle\citep{Ingram15}.

{\bf Specific questions for the next decade and beyond}

To improve our understanding of the formation and evolution of compact objects, we must (1) develop new surveys (2) both characterize the new objects detected in the surveys and re-characterize the objects in older surveys where there may be systematic uncertainties in the measurements.  At the present time we can just barely address questions like ``What are the typical masses of compact objects in close binaries?'' and ``Do a substantial fraction of black holes form with natal kicks?''  To move forward, we need to be able to address questions like ``Do the mass distributions of compact objects differ between the set in wide binaries and the set in close binaries?'' and ``Are the asymmetric natal kicks applied to black holes correlated with the masses of the black holes?''  The former question probes the effects of the common envelope process on the mass of the stellar remnant, while the latter can be used to determine whether, as is often suggested, there are multiple formation channels for black holes, with ``prompt collapse'' leading to higher mass black holes and no kicks, while black holes formed by fallback accretion onto neutron stars have smaller masses and asymmetric kicks generated during the temporary neutron star phase of their existence\citep{fryerkalogera01}.  

Additionally, it is further important information to understand whether the spins of the black holes correlate with the natal kick velocities or the black hole masses.  The processes for measuring spins are addressed in white papers by Javier Garcia and Jack Steiner.  We do note briefly that precise distance and inclination angle measurements are required to make use of the continuum fitting technique \citep{Steiner09}, while making use of models of the relativistic blurring of fluorescence lines gives the inclination angle as a fit parameter\citep{Garcia14} so that precise measurements of the inclination angles from other means can be used as checks on how well these models work.

The remaining discussions here will focus on binaries with little or no accretion and on binaries with low mass donor stars.  X-ray binaries with high mass donor stars are being discussed in at least one other white paper, by Nevin Vulic and collaborators. Furthermore, we focus on black holes because the mass and kick distributions of neutron stars are fairly well-established through use of pulsar measurements, and the primary motivation for new neutron star mass estimates is to understand the extrema to probe the neutron star equation of state.

{\bf New surveys}

Survey work is important both for establishing the overall sizes of the black hole populations, necessary for understanding the evolution of the binaries, and for collecting large samples of objects, necessary for being able to probe the properties of the compact objects and how they correlate with the properties of the binaries. Because about half of X-ray binaries are located in globular clusters it is important to study not just the field star populations\citep{kundu07}, but also to understand the dynamical processes that form close binaries with compact objects in the globular clusters, as well.   Because of the large number of parameters that can affect binary evolution, and because a wide variety of selection effects can be present in these surveys, it is important for a substantive modelling effort to be made simultaneously with the survey work to be able to interpret the surveys properly, including accounting for systems disrupted by supernovae and their associated natal kicks.

A wide range of techniques can be used to identify new compact objects to grow sample sizes for surveys.  At the present time, surveys for new accreting black holes  are done primarily in the X-ray band.  X-ray surveys will remain important in the future, but can be supplemented with surveys at optical and radio wavelengths.

In the X-ray band, X-ray binaries can be discovered from their outbursts.  Building up numbers of systems in this manner requires a high quality all-sky monitor.  Because a substantial fraction of X-ray binaries are embedded deep in the Galactic Plane, and because the faintest outbursts typically peak at hard X-rays, a monitor with substantial hard X-ray sensitivity, such as that for the Probe mission STROBE-X\citep{ray19}, is preferable.  The sensitivity of the monitors is especially important for discovering large numbers of short period X-ray binaries.\citep{Arur18}  

Quiescent black hole X-ray binaries will be most easily discovered in the radio bands.  Because the radio power of accretors scales as the X-ray luminosity to the 0.6-0.7 power\citep{Gallo03}, as black hole accretors get progressively fainter, it becomes relatively easier to detect them as radio sources.  Furthermore, with a facility like the Next Generation Very Large Array, with excellent sensitivity and angular resolution, it will be possible to make surveys of the Galactic Plane that have good enough sensitivity to provide the proper motion precision to astrometrically separate out background AGN from foreground sources.  This approach should be able to find about 100 new black hole X-ray binaries in a reasonable amount of ngVLA time.  These would include both low mass X-ray binaries over a wide range of orbital period, and some wide binaries which are weakly accreting wind-fed systems \citep{MaccaroneScienceBook}.

A few types of optical surveys will have sensitivity to new black hole binaries.  The very broad optical emission lines of X-ray binaries mean that they can be detected spectroscopically in ``blind'' surveys, or surveys which follow-up wide field surveys in the X-rays like eROSITA\citep{spiders}, and these broad lines can even be identified from carefully designed emission line filters\citep{Casares18}.  Additionally, a wide range of time domain optical surveys are already ongoing or planned for the near future, including LSST, ZTF, PanSTARRS, BlackGEM and ASAS, plus TESS.  These surveys all have the capability to find ellipoidally modulated binaries which can then be followed up further to establish their nature, and they can also find optical outbursts from X-ray binaries.  Ensuring that some high cadence data exist is of real importance both for using these surveys to find close binaries and to estimate the amplitudes of their ellipsoidal modulations.\citep{2018arXiv181112433S}

These approaches are all complementary.  X-rays will primarily be sensitive to sources in outburst, while radio will be sensitive to quiescent systems, but preferentially the brighter, and hence longer period quiescent systems\citep{garcia01}.  The optical emission line approach is most sensitive to the shortest period binaries, since these will have the broadest emission lines.  Optical spectroscopic monitoring searches for binaries and searches for ellipsoidal variables can find fully detatched black hole binaries\citep{Thompson18, Giesers18}.  Since the ngVLA will have the sensitivity to detect stars, both from their thermal emission\citep{CarilliScienceBook} and from their stellar winds\citep{MaccaroneWinds}, some astrometric wobble discoveries of black holes may be done with the ngVLA as well.  Astrometric wobble searches from {\it Gaia} should discover thousands of wide binaries with black holes \citep{Breivik17}.

{\bf System characterization}\\
{\it Distance and proper motion estimates}\\
Having precise distances is essential for making good estimates of both the masses and natal kicks of stellar mass black holes.  In the case of the mass, fixing the distance allows one to break degeneracies between orbital inclination angle and amount of accretion light.  Additionally, understanding the overall sizes of the accreting source populations requires an understanding of the volume over which they have been detected.

For black holes, precise geometric distances will most typically come from radio parallax measurements (since most X-ray binaries are strongly reddened meaning {\it Gaia}'s capabilities are relatively modest\citep{Gandhi19}), and the same measurements will also give precise transverse velocities.  With the Very Long Baseline Array, this technique has been established to work effectively, but the present time, it has been strongly sensitivity-limited, and only applied to a few sources.\citep{2014PASA...31...16M}  With a factor of 10 improvement in sensitivity relative to the VLBA, it will be possible to obtain geometric parallaxes for most known quiescent black hole X-ray binaries, and with higher frequencies of observation, it should be possible also to improve the precision of geometric parallax measurements in the radio proportionately.  The ngVLA would be a game-changer for this line of work.

{\it Mass estimates}\\
For sources which show astrometric wobble, mass estimates are relatively straightforward.  Otherwise, the classical techniques of measurements of radial velocity amplitudes, mass ratios and ellipsoidal modulations must be used.  The former two quantities can now be estimated through emission lines, using correlations based on basic disk structures and established to work empirically.\citep{Casares15,Casares16}.  Parallax distances aid in estimating the amplitudes of ellipsoidal modulations\citep{Cantrell10}. 

What is most important observationally for these techniques to be used is to have available large amounts of time on 4-8m class telescopes for the spectroscopic measurements, and to have the optical variability surveys devote substantial time to the Galactic Plane\citep{2018arXiv181112433S}, so that photometric monitoring apart from the large variability surveys is not necessary.  Follow-up of the globular cluster sources will require large telescopes which combine adaptive optics and spectroscopy.  It would be beneficial to bring MUSE-like capabilities into the US national observatory system.

\pagebreak
\subsection*{References}
{
\renewcommand{\section}[2]{}
\bibliographystyle{mn2e}
\bibliography{bibpops}
}

\end{document}